\documentclass[usenglish,notitlepage,a4paper,12pt]{article}
\usepackage{amsfonts}
\usepackage{amssymb}
\usepackage{eurosym}
\usepackage{float}
\usepackage{makeidx}
\usepackage{graphicx}
\usepackage{amsmath}
\usepackage{appendix}
\usepackage{scalefnt}
\usepackage{endnotes}
\usepackage{setspace}
\usepackage{lscape}
\usepackage{epstopdf}
\usepackage{xcolor}
\usepackage{epstopdf}
\usepackage{hyperref}
\usepackage{dcolumn}
\usepackage[margin=1in]{geometry}
\usepackage{natbib}
\usepackage{pdfpages}
\usepackage{booktabs}
\usepackage{caption}
\usepackage{epigraph}

\hypersetup{
    colorlinks = false,
    linkbordercolor = {green},
}

\makeatletter

\def\@biblabel#1{\hspace*{-\labelsep}}
\makeatother

\input{tcilatex}

\begin{document}

\title{{\LARGE The macroeconomic cost of climate volatility}\thanks{
We gratefully acknowledge Lars Peter Hansen, Paolo Barletta and seminar
participants at Banca d'Italia and the Qatar Centre for Global Banking \&
Finance at King's College London for their comments. We are thankful to
Fredrik Lindsten for providing example code on the particle Gibbs sampler
with particle rejuvenation. The views expressed in this paper are those of
the authors and do not necessarily reflect those of Banca d'Italia.}\\
\bigskip }
\author{{\large Piergiorgio~\textsc{Alessandri}}\thanks{%
Bank of Italy, Rome. E-mail: piergiorgio.alessandri@bancaditalia.it.} \\
{\ } \and {\large Haroon~\textsc{Mumtaz}}\thanks{%
Queen Mary University, City University of London, London. E-mail:
haroon.mumtaz@qmul.uk.}  \and  {\ }}
\date{February 2022}
\maketitle

\begin{abstract}
\noindent {\normalsize We study the impact of climate volatility on economic
growth exploiting data on 133 countries between 1960 and 2019. We show that
the conditional (\textit{ex ante}) volatility of annual temperatures
increased steadily over time, rendering climate conditions less predictable
across countries, with important implications for growth. Controlling for
concomitant changes in temperatures, a +1$^\text{o}$C increase in
temperature volatility causes on average a 0.3 per cent decline in GDP
growth and a 0.7 per cent increase in the volatility of GDP. Unlike changes
in average temperatures, changes in temperature volatility affect both rich
and poor countries.}

\bigskip

\textit{Keywords:}{\normalsize \ temperature volatility; economic growth;
panel VAR.}

\textit{JEL Classification:}{\normalsize \ C32, E32, Q54.}
\end{abstract}

\singlespacing

\bigskip \thispagestyle{empty}

\baselineskip=20pt\newpage



\setcounter{page}{1}

\epigraph{It is \textit{virtually certain} that hot extremes have become
more frequent and more intense across most land regions since the 1950s.} {%
IPCC Report n.6, August 2021}

\onehalfspacing

\section{Introduction}

\label{sec_intro}

Rising temperatures are known to have a negative impact on economic growth,
particularly in poor countries. In its last report, the Intergovernmental
Panel on Climate Change (IPCC) highlighted another important dimension of
climate change: fluctuations in climate conditions became larger over time,
with unprecedented swings in temperatures and precipitations affecting an
increasing number of geographical regions (\citealp{arias2021climate}). This
paper shows that, from an economic perspective, this phenomenon is as
important as the underlying change in temperature levels. 
We use climate data on 133 countries since the 1960s to estimate a panel VAR
with stochastic volatility. The model captures the endogenous interactions
between temperatures and economic activity and accomodates shocks that can
affect both the level and the variability of the underlying series. This
framework allows us to estimate the volatility of the residual change in
annual temperatures that cannot be forecasted using past data, quantifying
the \textit{ex-ante} `temperature risk' faced by households and firms in a
given country at a given point in time. Combined with appropriate
identification restrictions, it also allows us to isolate exogenous changes
in temperature volatilities and trace their impact on various economic
activity indicators. Our analysis yields two main results. The first one is
that temperature volatility increased steadily over time, even in regions
that were only marginally affected by global warming. The second one is that
temperature volatility matters for economic activity. Controlling for
temperature levels, a 1$^\text{o}$C increase in volatility causes on average
a 0.3 per cent decline in GDP growth and a 0.7 per cent increase in the
volatility of the GDP growth rate. In other words, volatile temperatures
lead at once to lower and more variable income growth. Volatility shocks
affect rich, non-agricultural countries too, and they are not driven by the
occurrence of large fluctuations in GDP, temperatures or precipitations. We
find that volatility impacts both consumption and investment, and that its
effects are larger for the manufacturing and services sectors. Our findings
demonstrate that risk plays an important role in the nexus between climate
and the economy. Economic agents respond to changes in the expected
variability of the environment, and, as in other macro-financial contexts,
lower predictability is by itself detrimental for growth. This suggests that
climate risk has important \textit{ex-ante} implications for welfare, and
that uncertainty on the future path of the climate system may affect the
economy before, and independently of, actual changes in realized
temperatures. 

\ 

\noindent \textbf{Related literature.} Our work lies at the intersection of
two strands of research. The first one studies the economic implications of
climate change. The negative influence of global warming on income and
welfare was originally highlighted using reduced-form Integrated Assessment
Models (IAMs), and more recently confirmed by general equilibrium models of
the interaction between climate and the economy. \footnote{%
See respectively \cite{tol2009economic}, \cite{stern2016economics} \cite%
{nordhaus2017survey} and \cite{Acemoglu2012TechnicalChange}, \cite%
{golosov2014optimal}, \cite{hassler2016environmental}.} A large body of
empirical evidence documents the relation between weather outcomes and
productivity, output and economic growth, as well as political stability,
migration or mortality (\citealp{dell2009temperature}; \citealp{dell2014we}%
). Although researchers broadly agree that rising temperatures reduce growth
in relatively poor countries, the evidence on developed economies is more
mixed (\citealp{burke2010democraticch}, \citealp{10.1257/mac.4.3.66}, %
\citealp{kahn2021long}). The ambiguity also arises in studies that focus on
agriculture: the negative influence of higher temperatures in EMEs is
uncontroversial (\citealp{dell2014we}), while studies based on
within-country variability in the USA reach conflicting conclusions (%
\citealp{deschenes2007economic}; \citealp{fisher2012economic}). 
We document a new ``volatility'' effect of climate change that operates over
and above the ``level'' effect studied in previous contributions. Our
results suggest that temperature volatility affects growth in developed
economies too. The second strand of research examines the macroeconomic
implications of changes in risk and uncertainty. The relevance of
macro-financial volatility for consumption, investment and production is
well documented in the literature.\footnote{%
See e.g. \cite{Bloom2009}, \cite{10.1257/aer.20131193}, \cite{CMR2014}, \cite%
{bksy2014}. Extensive surveys are provided by \cite{Bloom2014} and \cite%
{fernandez2020uncertainty}.} By studying the impact of temperature
volatility on growth we illustrate a new, thus-far ignored source of
aggregate risk for the business cycle. The existence of a time-varying
`climate risk' factor is consistent with recent evidence obtained from firms
and financial markets. Asset pricing models point to climate as an important
risk factor in the long run (\citealp{bansal2016price}), and suggest that
carbon risk and pollution are priced in the cross-section of stock returns (%
\citealp{bolton2020investors}; see also \cite{hong2020climate} and \cite%
{giglio2020climate}). 
Surveys and textual analyses of earning conference calls reveal that climate
risk considerations feature prominently in the decisions of institutional
investors (\citealp{krueger2020importance}) and listed firms around the
world (\citealp{sautner2020firm}; \citealp{li2020corporate}). Finally, local
temperature fluctuations can increase public attention to climate change and
push investors to tilt their portfolios towards low-emissions firms (%
\citealp{10.1093/rfs/hhz086}). 
Our work complements these studies by constructing empirical measures of 
\textit{ex-ante} temperature volatility for a large panel of countries,
documenting their historical patterns, and quantifying the macroeconomic
implications of exogenous volatility shocks.

\ 

We are aware of three existing studies of the linkages between climate
volatility and growth. \cite{donadelli2020computing} study the relation
between annual temperature volatility and output in England over the
1800-2015 period. \cite{kotz2021day} examine data on over one thousand
subnational regions over 40 years, showing that day-to-day temperature
variability reduces regional GDP growth rates. Using a spatial
first-difference design, \cite{linsenmeier2021} finds that day-to-day
variability (and to a lesser extent seasonal and interannual variability)
over the 1984-2014 period had a negative long-run effect on development,
proxied by nightlights observed in 2015. Since these works use \textit{%
realized} volatility or variability measures, the results lend themselves to
a number of different interpretations: higher volatility may imply that the
economy spends more time away from its optimal climate conditions, is hit
more frequently by extreme weather events, or is subject to higher
adaptation costs. 
Insofar as temperatures have nonlinear effects (%
\citealp{deschenes2011climate}; \citealp{burke2015global}; %
\citealp{barreca2016adapting}) and weather anomalies, hurricanes and
windstorms cause significant economic damages (\citealp{dell2014we}; %
\citealp{kim2021extreme}), realized volatility may capture the impact of
large first-moment shocks rather than the separate and potentially
independent role of second-moment shocks. We study instead \textit{%
conditional (ex-ante)} volatility measures that are conceptually different
from, and empirically unrelated to, large shocks and extreme events. Our
approach focuses on the role of risk and it aims to isolate the specific
influence of climate-related uncertainty on economic outcomes. 

\ 

The mechanisms through which \textit{ex ante} climate volatility could
affect the economy are conceptually similar to those that have been analyzed
in the literature on financial and macroeconomic uncertainty. If households
and firms adjust to changes in temperature levels, they are also likely to
respond to changes in volatility that shape the distribution of future
temperatures. In the online annex to the paper we resort to the dynamic
general equilibrium model developed by \cite{basu2017uncertainty} to explore
the nature and magnitude of this response. We study a simple extension of
the model in which temperatures follow an exogenous heteroscedastic process
and affect either the aggregate productivity of the economy or the rate at
which households discount future consumption streams. The simulations show
that, for plausible calibrations of the level effect of temperatures on
these fundamentals, temperature volatility becomes in itself a
quantitatively important driver of economic activity: a rise in volatility
has clear and broad-based recessionary implications, causing drops in
consumption, investment and prices that are comparable to those associated
with changes in temperature levels. Our simulations exploit simple shortcuts
to model the (as yet unknown) linkages between the economy and the climate
system, and we do not attempt a full structural estimation of the model.
However, the results consistently point to \textit{ex ante} climate risk as
an independent and quantitatively important driver of economic fluctuations.

\ 

The remainder of the paper is organized as follows. Section \ref%
{sec_model_frame} describes the data and introduces our empirical model, a
panel VAR with stochastic volatility. Section \ref{sec_results} illustrate
the main empirical results. Section \ref{sec_rob} explores various
robustness checks and extensions of the baseline model. In Section \ref%
{sec_mechanism} we examine the impact of volatility on investment,
consumption and sectoral value added measures. Section \ref{sec_con}
concludes the paper.



\section{Data and econometric methodology}

\label{sec_model_frame}

In our baseline analysis we rely on the climate dataset used in a widely
cited paper by \cite{10.1257/mac.4.3.66} (DJO). The dataset covers 133
countries and spans the years between 1961 and 2005. The main source is the
Terrestrial Air Temperature and Precipitation 1900-2006 Gridded Monthly Time
Series, which provides terrestrial monthly mean temperatures and
precipitations at 0.5x0.5 degree resolution. DJO aggregate the series to the
country-year level using population-weighted averages, with weights based on
population in 1990.\footnote{%
We refer the reader to DJO for details on the definitions of the variables
and the associated descriptive statistics.} A key advantage of this dataset
is that the estimates can be compared to those obtained in previous studies
on temperatures and growth, including DJO itself. In one the extensions of
Section \ref{sec_rob} we use as an alternative the Climatic Research Unit
gridded Time Series (CRU TS) dataset produced by the UK's National Centre
for Atmospheric Science at the University of East Anglia. This dataset
exploits newer and more accurate interpolation techniques and covers a
longer sample ending in 2019 (see \cite{harris2020version} for details).
Country-level observations are obtained in this case using area-weighted
rather than population-weighted averages, allowing us to examine the
robustness of the results along another potentially important dimension. All
macroeconomic series are sourced from the World Development Indicators
database maintained by the World Bank. The key variable in the baseline
analysis is real GDP per capita. In the supplementary regressions of Section %
\ref{sec_mechanism} we also use the total consumption to GDP ratio, the
investment to GDP ratio, where investment is defined as change in fixed
assets plus net change in inventories, and the annual growth rates of value
added in the agricultural, manufacturing and services sectors.

\ 

Our analysis has two related objectives. The first one is to estimate the
conditional volatility of annual temperatures for all countries in the
sample. These estimates provide a clear and rigorous measure of climate
predictability, and they allow us to assess whether predictability changed
at all since the 1960s. The second one is to test whether climate volatility
matters for economic growth, controlling of course for the influence of
global warming documented elsewhere in the literature. To achieve these
objectives in an internally-consistent fashion, we estimate a country-level
panel VAR model where (i) temperatures and economic growth are linked by a
two-way interaction; (ii) the residuals are heteroscedastic; and (iii)
changes in conditional volatilities can have first-order effects on
temperature and GDP growth. The following subsections describe structure,
identification and estimation of the model.\footnote{%
In treating the climate as a stochastic rather than a deterministic system
our framework is also consistent with recent views in climatology, see e.g. 
\cite{calel2020temperature}.}


\subsection{The panel VAR model}

The panel VAR model with stochastic volatility has the following form: 
\begin{equation}
Z_{it}=c_{i}+\tau _{t}+\sum_{j=1}^{P}\beta
_{j}Z_{it-j}+\sum_{k=0}^{K}\gamma_{k}\tilde{h}_{it-k}+v_{it}  \label{eq1}
\end{equation}

where $Z_{it}$ is a vector of endogenous variables and countries and years
are indexed by $i=1,2,...,M$ and $t=1,2,...,T$ respectively. The variance
covariance matrix $cov\left( v_{it}\right) =$ $\Omega _{it}$ is time-varying
and heterogenous across countries. This matrix is factored as:

\begin{equation}
\Omega _{it}=A^{-1}H_{it}A^{-1^{\prime }}  \label{eq2}
\end{equation}

where $A$ is a lower triangular matrix with ones on the main diagonal. $%
H_{it}$ is a diagonal matrix $H_{it}=diag\left( \exp (\tilde{h}_{it})
\right) $ where $\tilde{h}_{it}=[h_{1,it},h_{2,it},..h_{N,it}]$ denotes the
stochastic volatility of the orthogonalised shocks $\tilde{e}_{it}=Av_{it}$.
The stochastic volatilities follow a panel VAR(1) process:

\begin{equation}
\tilde{h}_{it}=\alpha _{i}+\theta \tilde{h}_{it-1}+b_{0}\tilde{\eta}_{it},%
\tilde{\eta}_{it}\sim N(0,1)  \label{eq3}
\end{equation}

where $\alpha _{i}$ denotes country fixed effects, $b_{0}$ is a lower
triangular matrix and $\tilde{\eta}_{it}=[\eta_{1,it}, \eta_{2,it},...,
\eta_{N,it}]$ denotes a vector of volatility shocks.

\ 

The distinguishing feature of the model is that volatilities appear as
regressors on the right-hand side of equation \ref{eq1}. Hence, if $%
\gamma_{k}\neq 0$, an exogenous increase in the volatility of any of the
variables included in the model can affect the dynamics of the entire
system. In our baseline specification we define $Z_{it}=\left[T_{it} \ \
\Delta GDP_{it}\right]^{\prime }$, where $T_{it}$ denotes average annual
temperature in degrees Celsius and $\Delta GDP_{it}$ is the annual growth
rate of real GDP. To make the notation more intuitive, in the remainder of
the paper we label the two level shocks ($e^T_{it}$, $e^{GDP}_{it}$), the
volatility processes ($h^T_{it}$, $h^{GDP}_{it}$) and the associated
volatility shocks ($\eta^T_{it}$, $\eta^{GDP}_{it}$). In this setup $%
e^T_{i,t}$ represents a shock to the temperature \textit{level} $T_{it}$; $%
h^T_{it}$ represents the log-volatility of $e^T_{it}$, i.e. of the
(residual) portion of $T_{it}$ that is unforecastable given past data; and $%
\eta^T_{it}$ represents a temperature \textit{volatility} shock, i.e. an
exogenous shift in volatility occurring in country $i$ at time $t$. We
interpret $h^T_{it}$ as an empirical measure of uncertainty about future
temperatures and $\eta^T_{it}$ as an unexpected temperature uncertainty
shock. \footnote{%
In the robustness analysis we consider a range of alternative specifications
that include e.g. annual precipitations or squared GDP and temperature
changes, and distinguish \textit{inter alia} between rich and poor countries.%
}

\ 

Intuitively, the model allows us to capture two mechanisms through which
temperature volatility could affect growth. The first one is the direct
impact of temperature volatility ($h^T_{it}$) on the growth rate of the
economy ($\Delta GDP_{i,t}$). Higher volatility may reduce foreign
investments, discourage risk-averse firms from undertaking new investment
plans, or force them to engage in costly adaptation and insurance programs.
The second one is a spillover from temperature volatility ($h^T_{it}$) to
output volatility ($h^{GDP}_{it}$). To the extent that changes in
temperatures matter for GDP growth rates, an increase in the frequency
and/or magnitude of those changes could render growth more volatile. 
\footnote{%
Notice that this channel is entirely independent of the first one. If
temperatures enter the production function, a volatility spillover could
arise even in a linear, risk-neutral and frictionless economy where $h^T$
has no direct influence on investment decisions.} The coexistence of these
mechanisms implies that climate uncertainty could affect welfare in two
ways, reducing an economy's average growth rate and rendering its behavior
more erratic over time.


\subsection{Identification and estimation}

The growth regressions traditionally employed in the literature treat the
climate system as an exogenous driving force. By contrast, our panel VAR
model allows for a two-way interaction between climate and economic
activity: in principle, $T_{it}$ can affect growth and respond endogenously
to changes in the level and volatility of $\Delta GDP_{it}$. As in any VAR
model, additional assumptions are thus needed in order to identify exogenous
temperature shocks. Our key identification assumption is that macroeconomic
developments have no contemporaneous (within-year) impact on climate
variables. We apply this assumption to both level and volatility shocks by
restricting the $A^{-1}$ and $b_0$ matrices to be lower-triangular. In
practice, this implies that $e^{GDP}_{it}$ and $\eta^{GDP}_{it}$ only affect 
$T_{it}$ and $h^T_{it}$ with a lag of at least one year. Recursive
identification schemes are notoriously problematic when dealing with
financial and macroeconomic data, but can be used safely in our application.
Development and technological change may alter temperature and precipitation
patterns over time, but this long-term phenomenon is very unlikely to have a
material impact over a one-year horizon. Even if it did, our approach would
approximate the data better than regression models that postulate strong(er)
forms of exogeneity of the climate indicators.

\ 

The model is estimated using a Gibbs sampling algorithm that is described in
detail in the technical appendix. The algorithm is an extension of methods
used for Bayesian VARs with stochastic volatility (see e.g. %
\citealp{clark2011real}) to a panel setting. The parameters of the model can
be collected into five blocks: $\left(\Gamma ,\bar{A},\bar{B},Q,\tilde{h}%
_{it}\right) $. Here $B=vec\left( \left[c_{i},\tau _{t},\beta _{1},..\beta
_{P},\gamma _{1},..\gamma _{K}\right] \right) $ denotes the coefficients of
equation 1, $\bar{A}$ is a vector that collects the elements of $A$ that are
not equal to $0$ or $1$, $\bar{B}=vec\left( \left[ \alpha _{i},\theta \right]
\right) $ while $Q=b_{0}b_{0}^{\prime }$ is the variance of the residual of
the transition equation (3). Each iteration of the algorithm samples from
the conditional posterior distributions of these parameter blocks. Given $%
\tilde{h}_{it}$ and $\bar{A}$, the model is simply a panel VAR with a known
form of heteroscedasticity. Therefore, given a normal prior, the conditional
posterior of $\Gamma $ is also normal after a GLS transformation. As
described in \cite{cogley2005drifts}, conditional on $\Gamma $ and $\tilde{h}%
_{it}$, the elements of $\bar{A}$ are coefficients in linear regressions
involving the residuals of the panel VAR. Therefore, their conditional
posterior is standard. Given $\tilde{h}_{it}$, equation (3) is simply a
panel VAR with fixed effects. As we employ conjugate priors for $\bar{B}$
and $Q$, their conditional posteriors are well known and easily sampled
form. With a draw of $\Gamma ,\bar{A},\bar{B},Q$ in hand, equations (1) to
(3) constitute non-linear state space model for each country. To draw from
the conditional posterior of $\tilde{h}_{it}$ we use the particle Gibbs
sampler of \cite{andrieu2010particle} and \cite{JMLR:v15:lindsten14a}, as
modified in \cite{2015arXiv150506356L} for state space models with
degenerate transitions. We use 55,000 iterations and retain every 10th draw
after a burn-in period of 5000 draws. In the technical appendix we show that
the estimated inefficiency factors are low, providing evidence in favor of
convergence of the algorithm.


\section{Empirical results \label{sec_results}}

The model in Section \ref{sec_model_frame} allows us to estimate the
conditional volatility of annual temperatures at the country level since the
1960s. These estimates offer a simple empirical characterization of
short-term `climate risk'. Conditional volatilities are intrinsically
forward-looking: they capture the magnitude of the fluctuations in
temperatures that are likely to materialize in each country at a given point
in time. Unlike realized volatilities, they are not mechanically driven by
the changes in temperatures observed in the recent past, including extreme
events. And, as long as the data is persistent, they convey information on
the likely evolution of the system: higher volatility signals the beginning
of a (potentially long) phase of erratic weather conditions. Hence, our
analysis captures a dimension of climate change and a transmission mechanism
that could, at least in principle, operate alongside the traditional 'level'
effect of rising temperatures documented elsewhere. In section \ref%
{sec_results_1} we discuss the evolution of temperature volatilities over
time and its relation to the global warming phenomenon discussed in the
literature. In section \ref{sec_results_2} we study the impact of exogenous
changes in temperature volatility on economic growth.


\subsection{Trends in temperature volatility\label{sec_results_1}}

\begin{figure}[h!]
\begin{center}
\includegraphics*[scale=0.55]{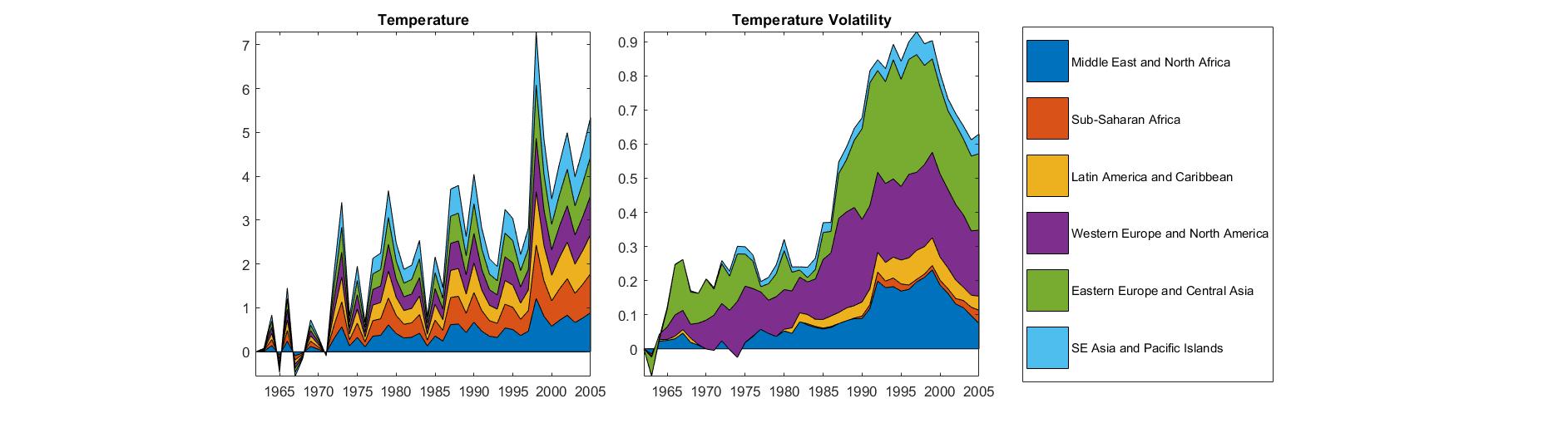}
\end{center}
\vspace{-3cm}
\caption{\textsc{{Trends in temperatures and temperature volatility}}} 
{\medskip	\noindent \small  The left panel shows the cumulative change in annual temperatures recorded between 1961 and 2005 in each of the six geographical regions listed in the legend. The right panel shows the cumulative change in each region's temperature volatility estimated by the panel VAR model. All figures are in degrees Celsius. The sample includes 133 countries and the regions are summarized by simple (unweighted) averages of country-level estimates.} \label{fig_01}
\end{figure}

Figure \ref{fig_01} shows the behavior of temperatures and conditional
temperature volatilities in six macro-regions between 1961 and 2005. Each
region is summarized by a simple (unweighted) average of its member
countries. The left panel replicates the stylized facts documented in DJO:
temperatures have risen across the board, particularly in the Middle East
and in Africa. The right panel shows that a similar trend occurred for
volatilities too. Volatility rose almost everywhere, with cumulative
increases of up to 0.2$^\text{o}$C in some of the regions. Shifts of this
magnitude could in principle have non-negligible economic implications. DJO
estimate that a 1$^\text{o}$C rise in annual temperatures reduces GDP growth
in poor countries by over 1 percentage point on average. \cite%
{BurkeEmerick2016} find that temperature changes of -0.5 to +1.5 $^\text{o}$%
C had a negative impact on agricultural output across US counties in the
past, suggesting that even rich and technologically sophisticated economies
may be vulnerable to climate fluctuations. The central question examined in
this paper is whether an increase in the likelihood of facing larger
temperature fluctuations in the future -- i.e. an increase in the
conditional volatility of annual temperatures -- can have similar effects on
growth (see Section \ref{sec_results_2}).

\ 

\begin{figure}[h!]
\begin{center}
\vspace{-8cm}
\includegraphics*[scale=.8]{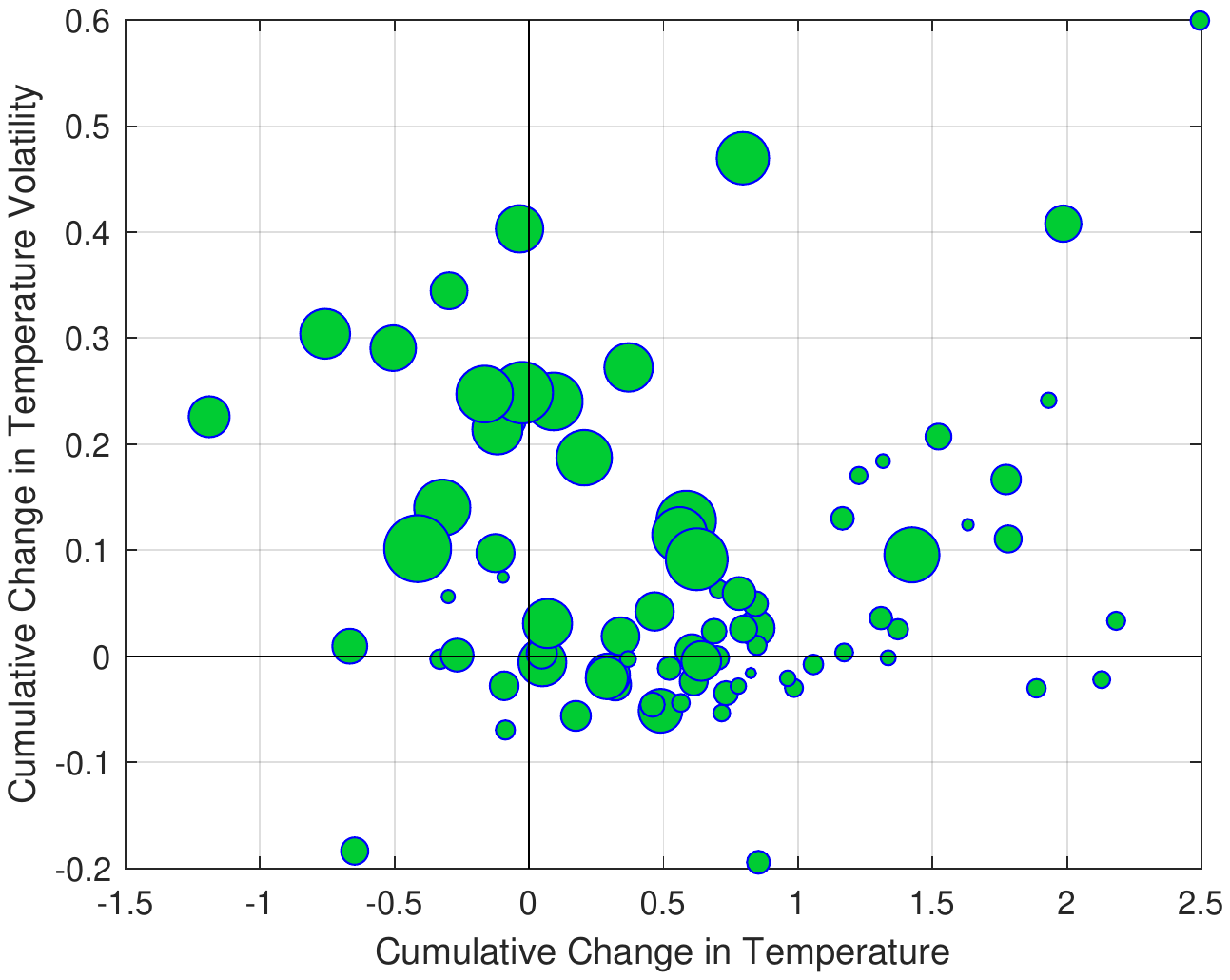}
\end{center}
\vspace{-8cm}
\caption{\textsc{{Correlation between temperature level and volatility}}} 
{\medskip	\noindent \small  Total change in annual temperatures (horizontal axis) versus total change in estimated temperature volatilities (vertical axis). Temperatures and volatilities are in degrees Celsius. The sample includes 133 countries between 1961 and 2005. The size of the bubbles represents the countries' average GDP levels in the 1950-1959 period. } \label{fig_02}
\end{figure}

The rise in volatility was relatively larger in Europe, Central Asia and
North America, which were only marginally affected by the increase in
temperature levels. This divergence is interesting \textit{per se} and it is
also useful for identification: if levels and volatilities were highly
correlated, it would be hard to disentangle their effects. To investigate
this point further, in Figure \ref{fig_02} we show the relation between
temperature levels and volatilities at the country level. The scatter plot
relates the cumulative change in temperatures recorded between 1961 and 2005
(horizontal axis) to the cumulative change in temperature volatility
estimated by the model (vertical axis). There is no correlation between
changes in levels and volatilities. In Section \ref{sec_results_2} we rely
on a more stringent identification strategy to estimate the impact of
volatility shocks; the model allows us to control for country and/or time
fixed effects as well as the lagged influence of GDP and temperatures.
However, the lack of correlation in figure Figure \ref{fig_02} yields
preliminary evidence that there is enough information in the data to
separate level and volatility effects. The scatter plot also shows that the
increase in volatility has been more widespread than the increase in average
temperatures. In particular, many large economies experienced a rise in
volatility combined with constant or decreasing average temperatures (see
north-western quadrant).\footnote{%
The scatterplot only includes 83 countries due to missing observations at
the beginning of the sample. The results are unchanged for the 1990-2005
window, which includes all 133 countries.}

\ 

Two further points are worth making. The change in volatilities over time is
highly significant from a statistical perspective. Figure A5 of the annex
shows the estimated average within-region volatility series together with a
68\% (one standard deviation) posterior coverage band. The null hypothesis
that volatility did not change between the 1960s and the early 2000s can be
safely rejected in all regions. In all but two cases, i.e. Sub-Saharan
Africa and Eastern Europe, volatility follows a clear upward trend at least
since the 1980s. At the same time, regional aggregations mask a significant
degree of heterogeneity at the country level. In figure A6 we compare the
confidence bands calculated at the regional level to the central estimates
of the country-specific volatilities. It is immediately clear that level,
variability and medium-run patterns in temperature volatility differ widely
across countries even within a given geographical region. 
The panel VAR allows us to exploit these forms of cross-country
heterogeneity to identify the causal effects of an exogenous change in
temperature volatility.


\subsection{The impact of temperature volatility on growth\label%
{sec_results_2}}

The panel VAR introduced in Section \ref{sec_model_frame} captures a number
of interactions involving both the level and the volatility of annual
temperatures and GDP growth at the country level. The posterior mean and
standard deviation of the parameters for the baseline model is summarized in
table A1 of the annex. The estimates highlight a potentially important
influence of volatility on both GDP and temperatures: all else equal, a rise
in temperature volatility $h^T$ is associated with lower growth and higher
temperatures. A rise in $h^{GDP}$ is also associated with lower growth
rates, which is consistent with the negative effects of economic volatility
on investment highlighted in the literature. Finally, $h^T$ and $h^{GDP}$
are both highly persistent, suggesting that the sample is characterized by
slow transitions between calm and volatile phases rather than sudden and
short-lived outbursts of volatility. 

\ 

\begin{figure}[h!]
\begin{center}
\includegraphics*[scale=.5]{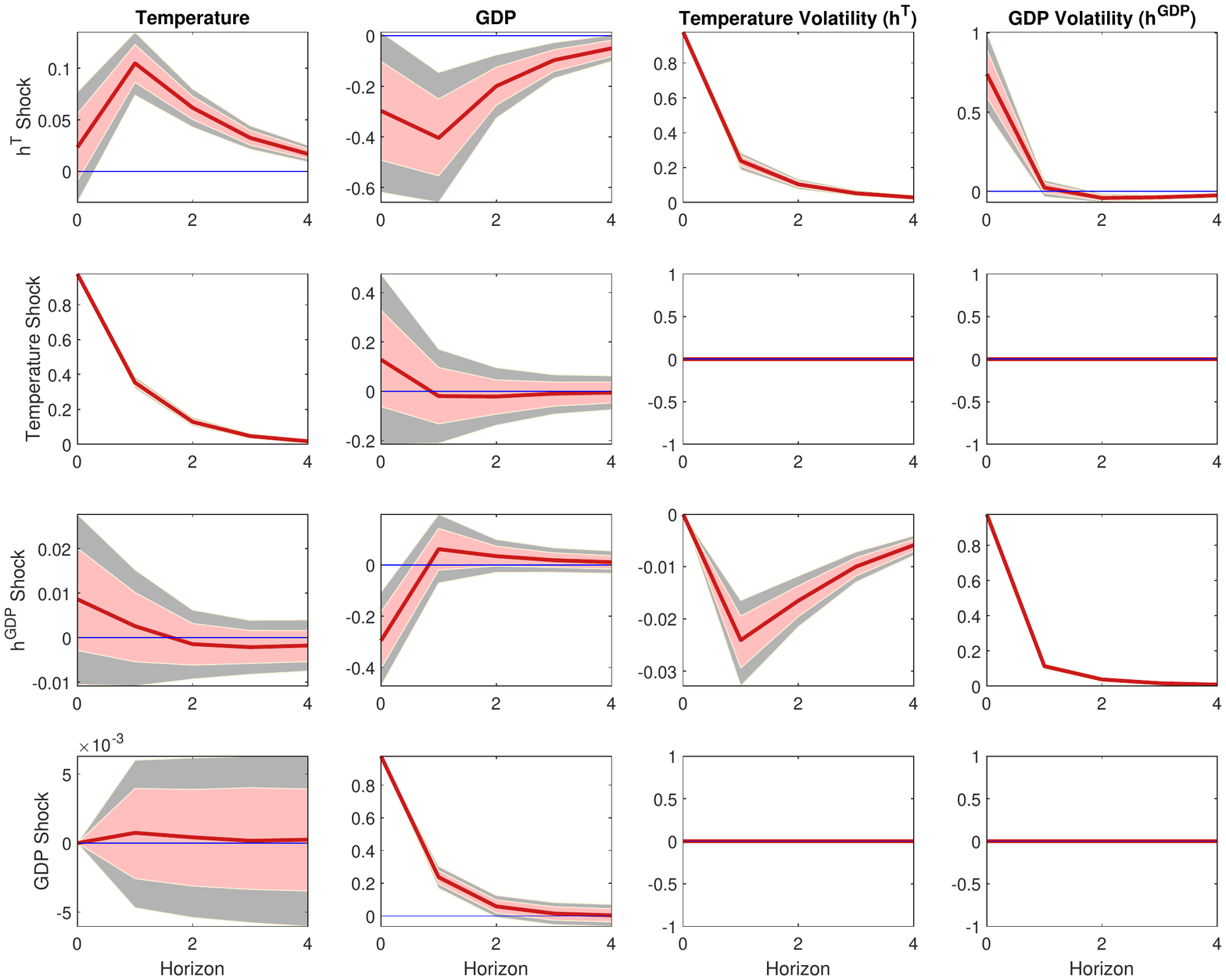}
\end{center}
\caption{\textsc{{Impulse-response functions}}} 
{\medskip	\noindent \small  Responses to level and volatility shocks in the baseline model. The estimates are obtained from a country-level panel VAR model where volatility is stochastic and changes in volatility influence the dynamics of all endogenous variables. \textit{Temperature} and \textit{GDP} are average annual temperature and annual GDP growth rate; $h^{T(GDP)}$ denotes the estimated conditional volatility of the temperature (GDP) series. The figure shows the responses to a 1$^\text{o}$C increase in $h^T$ (row 1), a 1$^\text{o}$C increase $Temperature$, a 1 percentage point increase in $h^{GDP}$ (row 3) and a 1 percentage point increase in $GDP$ (row 4). In all cases we report the mean responses with 68\% and 90\% posterior coverage bands. The estimation sample includes 133 countries between 1961 and 2005. } \label{fig_03}
\end{figure}

In Figure \ref{fig_03} we report the impulse-response functions associated
to the four shocks included in the model. For each variable the figure
reports the estimated mean response with 90\% and 68\% posterior coverage
bands. The first row of the figure shows the impact of an exogenous +1$^%
\text{o}$C increase in temperature volatility ($h^T$, column 3). The shock
causes an increase in temperatures, a decline in GDP growth and an increase
in the volatility of the GDP process. The responses are statistically
significant and fairly large in quantitative terms: GDP drops by 0.3 percent
on impact and remains below equilibrium for four years (col. 2), while the
conditional volatility of the GDP growth rate rises by about 0.75 percentage
points (col. 4). The results imply that countries that experience high
temperature volatility in a given year are likely to grow less and face more
pronounced GDP fluctuations in the medium term. In other words, higher
temperature volatility brings along a weaker and riskier growth path. It is
worth emphasizing that the transmission mechanism hinges on temperature risk
rather than actual changes in temperature levels. First, $h^T_{i,t}$
measures by construction the conditional volatility of temperatures in
country $i$, and it is unrelated to the weather patterns observed in $i$ in
the past. Second, the model controls for the influence of both
contemporaneous and past temperature shocks on GDP growth. Third, the
estimates remain unaltered if we include squared annual changes in
temperatures or precipitations to control for the confounding influence of
`extreme' weather events (see Section \ref{sec_rob}). Hence, the results
show that economic agents respond to the degree of expected variability of
the environment, and that -- as in the case of many other non-climatic
factors -- lower predictability is \textit{per se} detrimental for growth.

\ 

Figure \ref{fig_03} shows that temperature and output shocks have negligible
implications for the system, except for raising respectively $T_t$ and $%
\Delta GDP_t$ (rows 2 and 4). The impact of a temperature increase on GDP is
positive but non-significant in the baseline specification, but it becomes
negative if the model is estimated over poor countries only (see Section \ref%
{sec_rob}). Both results are in line with those reported in DJO. Interesting
results emerge for shocks to the volatility of the GDP process ($h^{GDP}$,
row 3). A rise in GDP volatility has a negative impact on the GDP growth
rate, consistent with the well-known influence of macroeconomic uncertainty
on economic activity (see e.g. \cite{Bloom2014} and \cite%
{fernandez2020uncertainty}). However, the impact is smaller and extremely
short-lived: instead of a protracted slowdown, the shock causes a one-year
drop followed by a quick rebound with mild signs of `overshooting'. This has
important implications for the intepretation of the results in the first row
of Figure \ref{fig_03}. On the one hand, endogenous increases in
macroeconomic volatility contribute to the propagation of temperature
volatility shocks. One of the reasons why temperature volatility reduces
economic activity is that it makes the growth rate of the economy less
predictable. On the other hand, this volatility spillover must be part of a
more complex transmission mechanism. The change in macroeconomic volatility 
\textit{per se} does not account for the decline in GDP, particularly over
longer horizons, suggesting that temperature volatility has an additional,
direct effect on the economy. We investigate the nature of the transmission
mechanism more closely in Section \ref{sec_mechanism}.

\ 

The analysis carried out in this Section leads to two conclusions. First,
temperature volatility has risen steadily across countries since the 1960s,
rendering climate conditions less predictable over time. Second, controlling
for temperature levels, a rise in temperature volatility is followed by a
period of low and volatile GDP growth rates. Taken together, these findings
point to two distinct mechanisms through which climate change could affect
income and welfare over and above the well-known `global warming'
phenomenon. In the next section we examine the robustness of the baseline
results along various dimensions, considering \textit{inter alia} the role
of precipitations, cross-country heterogeneity, the relation between
volatility and extreme events, and the alternative CRU-TS climate dataset.


\section{Robustness and extensions}

\label{sec_rob}

In this section we replicate the baseline analysis using alternative
specifications and estimation samples for the panel VAR model. To save space
we only discuss the impact of temperature volatility shocks on level and
conditional volatility of the GDP growth rate; the online annex provides
more details. The main results of the tests are summarized in figures \ref%
{fig_04} to \ref{fig_06} and in tables \ref{table_shortrun} and \ref%
{table_longrun}. The figures compare the impulse-response functions from the
alternative specifications to those obtained in the baseline case. The
tables report the central estimates of the short-run and long-run impacts of
the shock on GDP in each model, along with the corresponding 68\% posterior
coverage intervals.


\subsection{Heterogeneity}

\label{sec_rob_heterogeneity}

\begin{figure}[h!]
\begin{center}
\vspace{-2cm}
\includegraphics*[scale=.6]{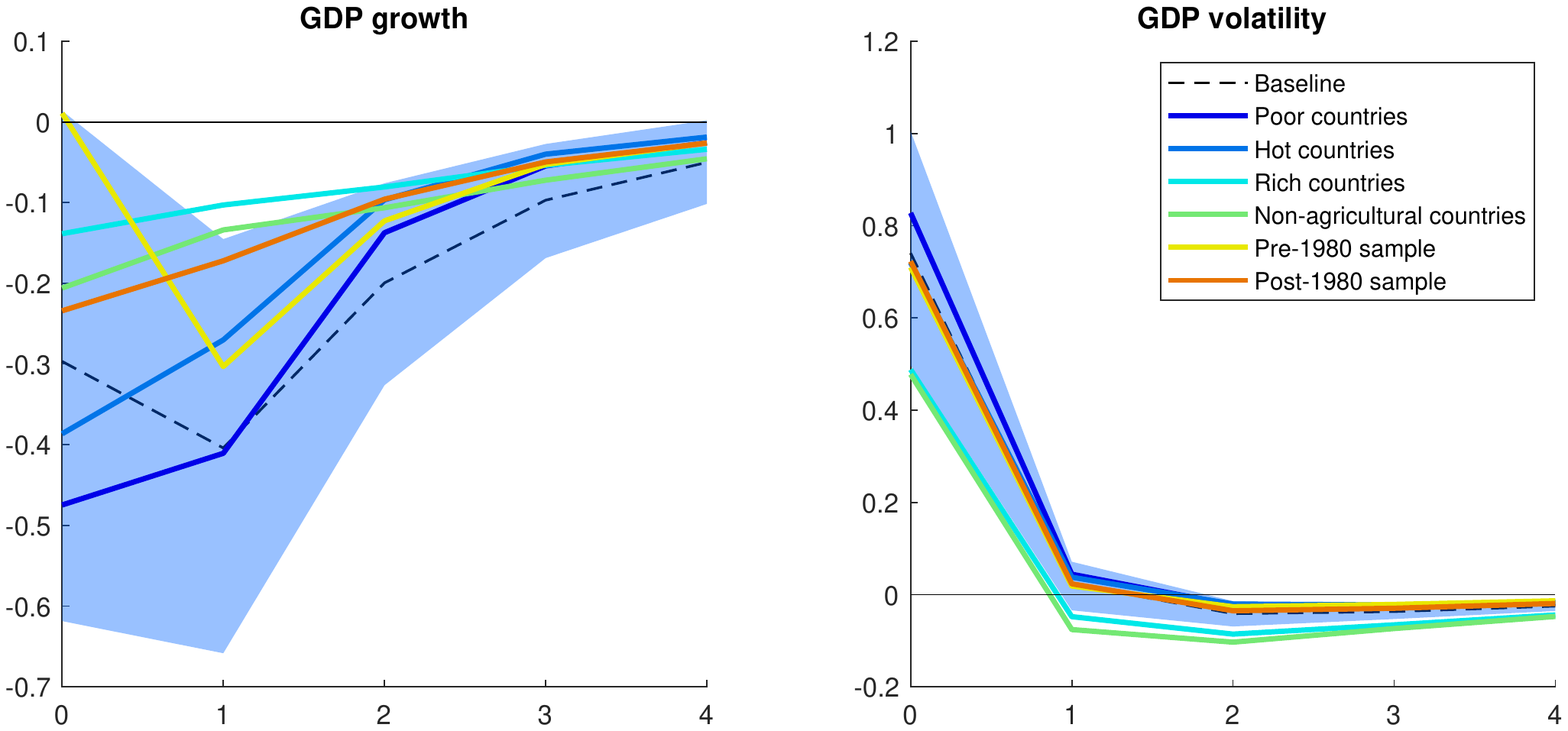}
\end{center}
\vspace{-4cm}
\caption{\textsc{{Heterogeneity}}} 
{\medskip	\noindent \small  Impact of a 1$^\text{o}$C increase in temperature volatility on the annual growth rate of GDP (left panel) and its conditional volatility (right panel). The shaded area is the 90\% posterior coverage band obtained from the baseline specification of the panel VAR model. The additional lines represent the central estimates obtained restricting the estimation to a subsample of poor, hot, rich or non-agricoltural countries, or splitting the sample in 1980.} 
\label{fig_04}
\end{figure}

There is an open debate on the cross-sectional and distributional
implications of climate change. In particular, rising temperatures may
affect only or mostly poor countries that rely heavily on agricolture and
have limited adaptation capabilities (see Section \ref{sec_intro}). The
first issue we investigate is thus how the macroeconomic effects of climate
volatility vary across regions. We re-estimate the baseline panel VAR
specifications using only data on ``poor'', ``rich'', ``hot'', or
``non-agricultural" countries. All groups are defined using the dummy
variables suggested by DJO. Poor is a dummy for countries that have
below-median GDP per capita in their first year in the dataset, hot is dummy
for countries with above-median average temperature in the 1950s, and
agricultural is a dummy for countries with an above-median share of GDP in
agriculture in 1995. The estimates are reported in Figure \ref{fig_04},
which compares the impact on GDP of a one-standard deviation increase in
temperature volatility obtained in the four alternative subsamples. The
responses are qualitatively similar across samples: in all cases GDP growth
is lower and more volatile on impact, i.e. in the year when volatility
rises, and for up to four years after the shock. The differences across
subsamples largely mirror those documented for temperature levels: rich and
non-agricultural countries are less affected than poor or hot countries.
However, heterogeneity is less pronounced. On impact, the shock raises GDP
volatility in all groups of countries, but it causes a statistically
significant GDP contraction only in poor and hot countries (see table \ref%
{table_shortrun}). At the five-year horizon the situation is reversed: the
cumulative GDP response becomes significant for all groups, while the
volatility response is only significant for poor and hot countries (see
table \ref{table_longrun}). This suggests that rich countries smooth the
impact of the shock rather than averting it altogether. They insulate GDP
from shocks that take place within a given year, and succeed in mitigating
the macroeconomic uncertainty caused by those shock, but they cannot avoid
the longer-term implications of higher temperature volatility. The estimates
GDP loss caused by a +1$^\text{o}$C increase in temperature volatility is
-0.48\% for rich countries and -1.14\% for poor countries. All in all, the
evidence indicates that climate \textit{volatility} matters even for
highly-developed economies that can adjust efficiently to gradual changes in
temperature \textit{levels}.

The last two lines in Figure \ref{fig_04} show the response obtained
estimating the model separately on pre- or post-1980 observations. The GDP
response is slower in the earlier sample, but the pattern of the
impulse-response functions is otherwise fairly similar. Tables \ref%
{table_shortrun} and \ref{table_longrun} confirm that, although the
contemporaneous GDP response is not significant before 1980, the long-term
response is significant and almost identical in the two samples (-0.5\% 
\textit{versus} -0.6\%).


\subsection{Alternative specifications}

\label{sec_rob_specifications}

\begin{figure}[h!]
\begin{center}
\vspace{-2cm}
\includegraphics*[scale=.6]{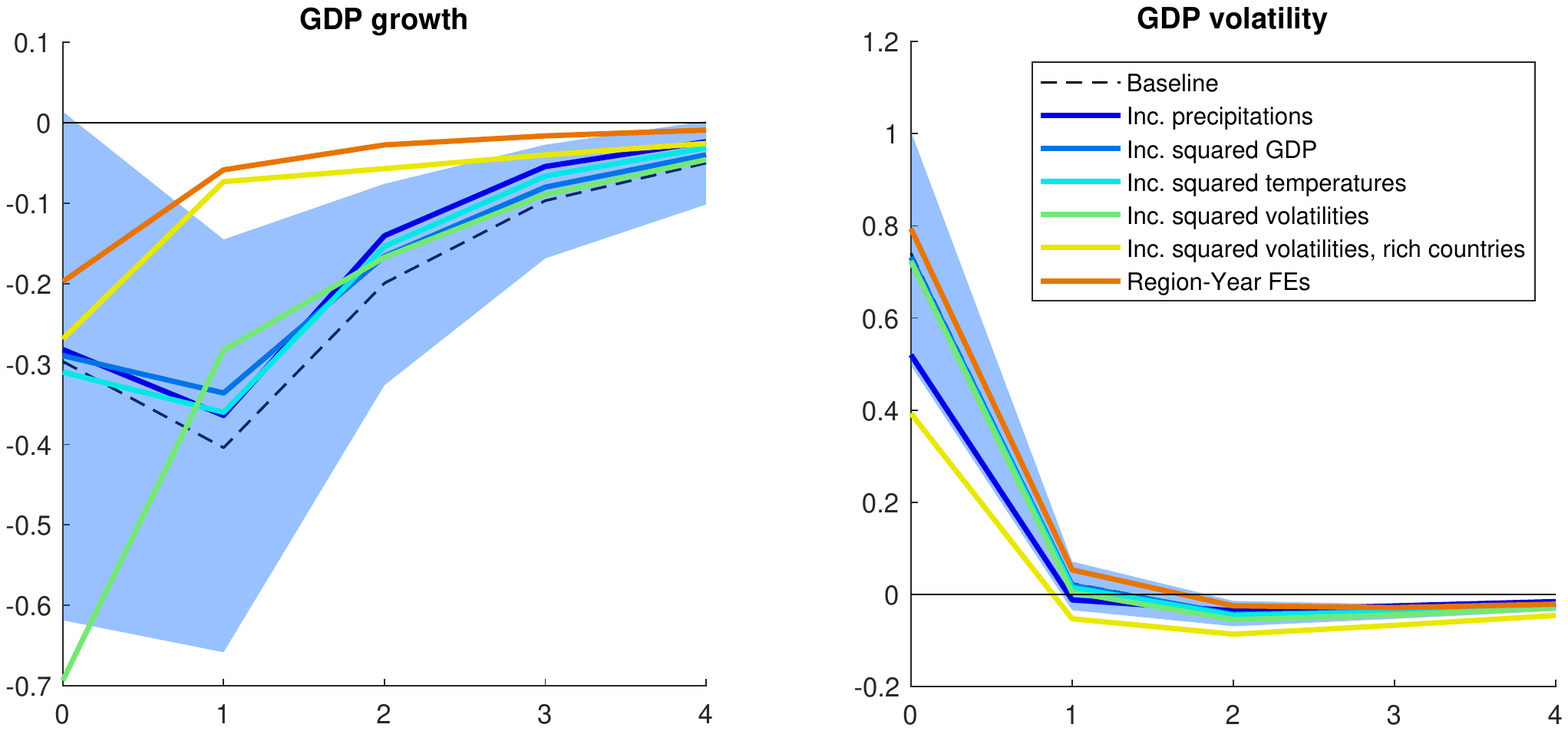}
\end{center}
\vspace{-4cm}
\caption{\textsc{{Additional controls}}} 
{\medskip	\noindent \small  Impact of a 1$^\text{o}$C increase in temperature volatility on the annual growth rate of GDP (left panel) and its conditional volatility (right panel). The shaded area is the 90\% posterior coverage band obtained from the baseline specification of the panel VAR model. The additional lines represent the central estimates obtained by alternatively including in the model average yearly precipitations, squared GDP growth rates, squared temperatures, squared GDP and temperature volatilities, or a set of region-by-year fixed effects.} 
\label{fig_05}
\end{figure}

\begin{figure}[h!]
\begin{center}
\vspace{-7cm}
\includegraphics*[scale=.80]{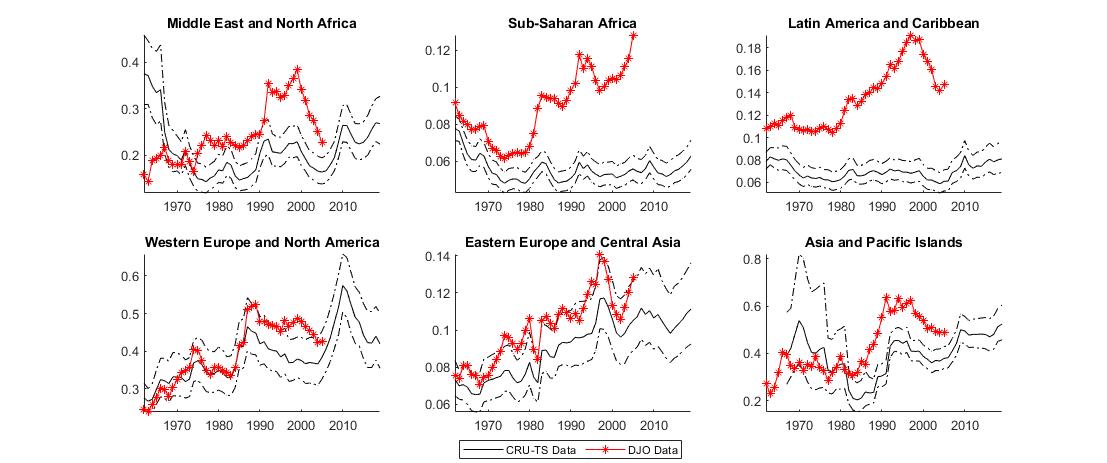}
\end{center}
\vspace{-7cm}
\caption{\textsc{{Temperature volatility in the CRU-TS dataset}}} 
{\medskip	\noindent \small  The figure compares the conditional volatility of annual temperatures in the Climatic Research Unit gridded Time Series dataset (CRU-TS, in black) to those in the \cite{10.1257/mac.4.3.66} dataset (DJO, in red). For the CRU-TS data we report the 16th and 84th percentiles of the posterior distribution along with the median estimates. All series are obtained using the baseline specification of the panel VAR model and are expressed in Celsius degrees. See notes to figure \ref{fig_01}.} 
\label{fig_06}
\end{figure}

As a next step, we extend the baseline model to account for other factors
that might affect the climate-growth nexus. The results of these tests are
displayed in figure \ref{fig_05}. We first replicate the baseline analysis
adding to the model precipitations ($P_{it}$, in units of 100 mm per year),
which are routinely employed together with temperatures as a proxy of
climatic change. The identification of the shocks is again based on a
recursive ordering of the variables. We place precipitations before GDP so
to maintain the assumption that the climate is exogenous to macroeconomic
shocks in the short term (see Section \ref{sec_model_frame}). This model
delivers slightly smaller estimates of the peak impact of temperature
volatility shocks on GDP and GDP volatility, but the responses remain
statistically significant (see table \ref{table_shortrun}). \footnote{%
Shocks to the volatility of temperatures and precipitations (i.e. to $%
h^T_{it}$ and $h^P_{it}$ ) have a qualitatively similar influence on GDP,
which suggests that the 'volatility channel' operates through both
temperatures and precipitations.}

\ 

A particularly critical issue is the potential nonlinearity of the link
between climate and the economy. 
Output and productivity decline globally when temperatures move
significantly below or above 13$^\text{o}$C (\citealp{burke2015global}).
Mortality rates rise sharply in the US in areas and periods in which
temperatures reach the upper percentiles of their distribution (%
\citealp{deschenes2011climate}; \citealp{barreca2016adapting}). Furthermore,
weather anomalies are known to have a large negative impact on the economy (%
\citealp{dell2014we}; \citealp{kim2021extreme}). This creates a non-trivial
identification challenge: from an empirical perspective, a volatility proxy
could simply capture the nonlinear impact of large shocks to temperature
levels. In previous studies on the relation between climate volatility and
growth, \cite{donadelli2020computing}, \cite{kotz2021day} and \cite%
{linsenmeier2021} employed realized \textit{(ex-post)} volatility or
variability measures that are by construction affected by the occurrence of
large fluctuations in weather conditions. \cite{donadelli2020computing}
report a correlation between temperature volatility and extreme events --
defined as heavy rainfalls, floods, frosts, hot temperature anomalies and
droughts -- of 0.59 in the UK in the post-war period (see figure 1 of the
paper). \cite{kotz2021day} obtain an annual variability measure by
calculating the intra-monthly standard deviation of daily temperatures and
then averaging it over 12 months. The averaging step ameliorates the problem
but it is unlikely to solve it completely: we find that in the pooled
dataset the indicator has a correlation of 0.30 with the squared change in
annual temperatures 
In \cite{linsenmeier2021} day-to-day variability (which is heavily affected
by large shocks) has indeed a stronger economic impact than seasonal or
interannual variability (which are smoother and less affected by those
shocks due to temporal aggregation). The upshot is that the relation between
realized volatility and GDP may mask the impact of nonlinearities and large
shocks rather than a genuine uncertainty component. 

Our \textit{ex-ante} volatility measures are not subject to this limitation
because they are not constructed using (and are thus unrelated to) past
temperatures. However, the identification problem may in principle arise in
our case as well. 
As a first check we examine the correlation between changes in the estimated
conditional temperature volatilities ($h^T_{i,t}$) and squared annual
changes in temperatures, precipitations and income ($T_{i,t}$, $P_{i,t}$, $%
GDP_{i,t}$). These provide rough estimates of the realized annual volatility
of the three series, capturing a range of 'extreme events' -- i.e. large
year-to-year shifts in temperatures, precipitations or GDP -- that may
potentially bias the estimation of the climate risk effect. The correlations
are extremely low for all geographical regions, which means that
fluctuations in $h^T_{i,t}$ do not systematically overlap with large
year-on-year changes in output or weather conditions (see figure 7 in the
annex). We then re-estimate the panel VAR adding to the baseline variables
squared GDP growth rates, squared temperatures, or the squared volatility
terms. As figure \ref{fig_05} shows, none of these changes has major
implications for our results. The contemporaneous impact of the shock on GDP
growth and GDP volatility is virtually unaffected by the inclusion of
squared GDP and temperatures, and it almost doubles when squared
volatilities are added to the model (see \ref{table_shortrun}). Furthermore,
the long-run impact of the shock on GDP varies between -0.9\% and -1.3\%,
close to the baseline estimate of -1.1\% (see table \ref{table_longrun}).
The tables show that, in the case of rich countries, both the short-term and
the long-term impact of the shock become in fact larger and more significant
when the squared volatility terms are added to the model. These results
corroborate the conclusion that the model picks up the specific influence of
the conditional volatility of annual temperatures rather than nonlinearities
involving past GDP and/or realized temperature fluctuations.

\ 

As a final test, we estimate the baseline model including a set of year$%
\times$region fixed effects to control for the potential influence of
unobserved and time-varying drivers of GDP growth. The GDP volatility
response is virtually unaffected by the change, while the GDP response
becomes smaller and less persistent (see figure \ref{fig_05}). The estimated
short- and long-run GDP responses drop respectively to -0.2\% and -0.3\%.
Both are statistically significant. This confirms that, although the
baseline estimates may be somewhat distorted by specific subperiods or
regional trends, temperature volatility has a specific well-defined
influence on economic growth.


\subsection{An alternative climate dataset}

\label{sec_rob_newdata} The DJO dataset facilitates a comparison between our
results and those available in the literature, but it ends in 2005. Below we
replicate the estimation of the baseline model using the Climatic Research
Unit gridded Time Series (CRU TS) dataset produced by the UK's National
Centre for Atmospheric Science, that covers a longer sample ending in 2019.
The data is described in detail in \cite{harris2020version}. CRU-TS relies
on more recent techniques to interpolate station-level weather observations;
it also employs area-weighted rather than population-weighted averages to
construct country-level observations, allowing us to examine the robustness
of our conclusions along another important dimension. Figure \ref{fig_06}
compares the temperature volatility series obtained by estimating the
baseline panel VAR model on the two datasets.\footnote{%
We focus on the six geographical regions that appear in figure \ref{fig_01}.
A more detailed description of the CRU-TS estimates is available in figures
8 and 9 of the annex.} The estimates follow fairly similar patterns over the
1960-2005 period in four of the six regions. Temperature volatility rose
markedly in Europe, North America and Asia (bottom row) and remained roughly
constant in the Middle East and North Africa (top left panel). The magnitude
of the rise in volatility is also generally similar; for Western Europe and
North America, for instance, volatility is estimated to rise from about 0.2$^%
\text{o}$C in 1965 to 0.4$^\text{o}$C in 2005. Larger differences appear in
the case of Sub-Saharan Africa and Latin America, for which volatilities
rise in the DJO data and remain constant in the CRU-TS data. The presence of
large and sparsely populated countries makes the weighting scheme more
influential in these regions. The CRU-TS data also shows that volatility
rose or remained constant from 2005 onwards. Tables \ref{table_shortrun} and %
\ref{table_longrun} show that temperature volatility shocks have similar
implications in the two datasets. The estimated impact of the shocks is
actually larger and statistically more significant in the CRU-TS dataset: a
+1$^\text{o}$C increase in volatility causes GDP to drop by 0.4\% on impact
and 1.3\% in the long run. The conditional volatility of GDP also rises more
compared to the baseline.\footnote{%
A full set of impulse-reponses is provided in figure 10 of the annex.} 


\section{Transmission mechanisms}

\label{sec_mechanism}

Pinning down the mechanisms through which climate shocks affect the economy
is difficult because these shocks influence a broad range of outcomes,
including \textit{inter alia} output, productivity, health or migrations (%
\citealp{dell2014we}). Climate conditions affect aggregate supply directly,
through their impact on infrastructures, natural resources and trade flows;
but they can also shift aggregate demand by influencing households' income,
wealth and consumption patterns (\citealp{batten2020climate}). The
transmission mechanisms are also likely to vary across countries. Natural
disasters have a null or mildly negative impact on inflation in rich
countries, but cause long-lasting price increases in emerging countries (%
\citealp{parker2018impact}). Among OECD countries, spikes in green-house gas
emissions trigger joint declines in output and prices that mimic a weakening
in aggregate demand (\citealp{ciccarelli2021demand}). In short, there seems
to be no single dominant transmission mechanism for the `level' shocks
traditionally studied in the climate literature. Our simulations based on
the \cite{basu2017uncertainty} model suggest that the propagation of
volatility shocks is equally complex: a rise in volatility causes a slowdown
under a broad range of model configurations, but the relative movements in
consumption, investment and prices vary depending on the mechanism that
links temperatures to the fundamentals of the economy (see section A of the
online annex).

\ 

\begin{figure}[h!]
\begin{center}
\vspace{-6cm}
\includegraphics*[scale=.75]{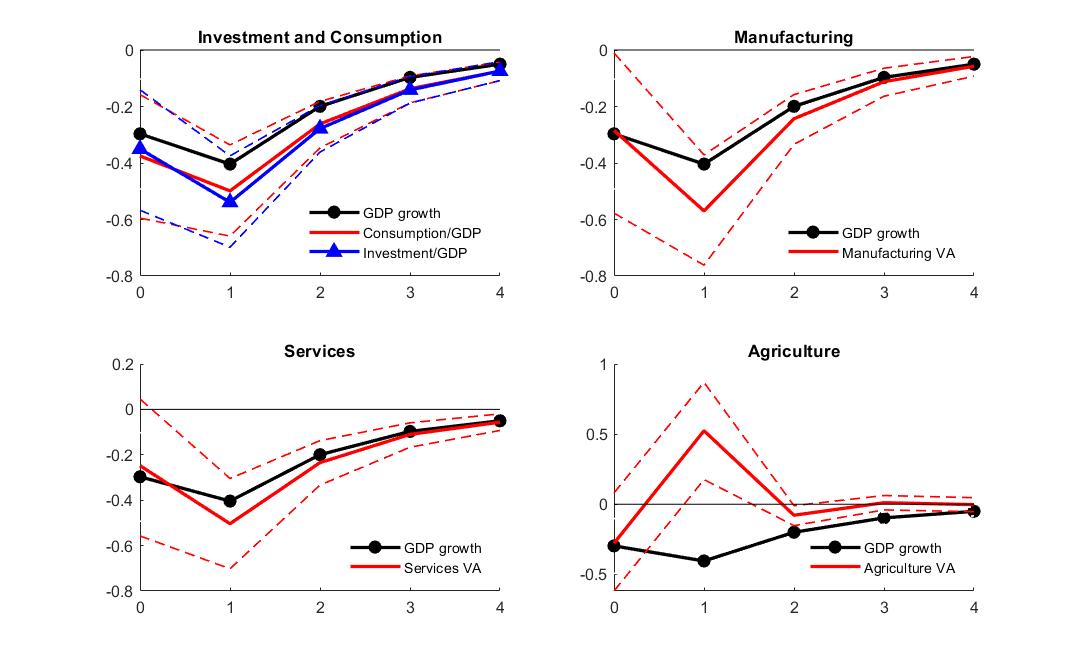}
\end{center}
\vspace{-7cm}
\caption{\textsc{{Transmission of temperature volatility shocks}}} 
{\medskip	\noindent \small  The figure shows the impact of a  1$^\text{o}$C increase in temperature volatility on the consumption-to-GDP and investment-to-GDP ratios (top left panel) and on the annual growth rates of value added in the manufacturing, services and agricultural sectors (top right and bottom panels). All panels include for comparison the estimated impact of the shock on annual GDP growth derived from the baseline panel VAR model. The estimation sample includes 133 countries between 1961 and 2005.} 
\label{fig_07}
\end{figure}

To shed light on the transmission mechanisms we estimate an additional set
of panel VAR specifications in which GDP growth is replaced by alternative,
more disaggregated indicators of economic activity. We first consider
investment (defined as gross capital formation) and total consumption
expenditure, both expressed as percentages of GDP. The top left panel of
Figure \ref{fig_07} shows the responses of these variables to the 1$^\text{o}
$C increase in temperature volatility examined in the previous section. The
GDP response obtained from the baseline model is also shown for comparison.
Investment and consumption drop by a maximum of about 0.5 percentage points,
following a trajectory that closely follows that of GDP. 
We next examine the annual growth rate of value added (VA) in the
manufacturing, services and agricultural sector. VA captures changes in
productivity and in the price and/or mix of production factors employed in
each sector. A rise in volatility has a persistent negative impact on
manufactures and services, consistent with a generalized drop in
productivity or a less intense utilization of capital. The shock has instead
a positive but short-lived impact on agriculture; one possibility is that VA
is driven in this case by a relatively large increase in the price of the
final goods (see \citealp{loayza2012natural}). 
The estimates in figure \ref{fig_07} paint a fairly clear picture. An
exogenous rise in temperature volatility causes a drop in consumption and
fixed investment that is consistent with the precautionary response and the
wait-and-see effect that are generally associated to a rise in economic or
financial uncertainty. Furthermore, volatility shocks, like changes in
realized temperatures, cause a slowdown that spreads widely across the main
sectors of the economy. We leave a more granular,
geographically-differentiated investigation of the transmission mechanisms
to future work.


\section{Conclusions}

\label{sec_con}

Rising temperatures are known to have a negative impact on economic growth,
particularly in poor countries. This paper shows that climate change also
affects economic outcomes through a volatility channel. We use a panel VAR
model with stochastic volatility to identify exogenous changes in
temperature volatility and assess their implications for the macroeconomy.
We exploit the model to estimate the conditional volatility of annual
temperatures for 133 countries between 1961 and 2019. These estimates
capture the variability of the residual component of annual temperatures
that cannot be predicted using past data, quantifying the \textit{ex-ante}
`temperature risk' faced by households and firms in a given country at a
given point in time. The model captures the interaction between levels and
variances of annual temperatures and GDP growth rates, allowing the
identification of exogenous temperature volatility shocks. The analysis
yields two main conclusions. First, temperature volatility increased
steadily over time, even in regions that were only marginally affected by
global warming. Second, temperature volatility matters for growth. Changes
in volatility affect both the means and the variances of the GDP growth
rates of the countries in our sample. Controlling for temperature levels, a
+1$^\text{o}$C increase in volatility causes on average a 0.3 per cent
decline in GDP growth and a 0.7 per cent increase in the volatility of the
GDP growth rate. These mechanisms operate in rich, non-agricultural
countries too, and they are statistically and economically significant even
controlling for the influence of large realized fluctuations in GDP,
temperatures or precipitations. Our findings demonstrate that economic
agents respond to changes in the expected variability of the environment.
They also suggests that climate risk may have important \textit{ex-ante}
implications for welfare, as uncertainty has economic costs that can
materialize before, and independently of, any observed change in
temperatures.

\newpage

\singlespacing

\bibliographystyle{CHICAGO}
\bibliography{references,RefFile_v1}


%
%
%
%
%


\clearpage \pagebreak 

\begin{table}[h!]
\begin{tabular}{lrrrrrrr}
\toprule
                                      & \multicolumn{3}{c}{GDP} 		& 			&  \multicolumn{3}{c}{GDP volatility} \\
                                      & 16\%                 & 50\%                           & 84\%                 &       \ \ \		& 16\%                 & 50\%                               & 84\%                 \\
\midrule
Baseline                                    & -0.492               & -0.297               & -0.102               &                   & 0.597 & 0.741 & 0.896 \\
Poor countries                              & -0.772               & -0.475               & -0.162               &                      & 0.624 & 0.828 & 1.024 \\
Hot countries                               & -0.663               & -0.386               & -0.121               &                      & 0.543 & 0.720 & 0.903 \\
Rich countries                              & -0.308               & -0.138               & 0.028                &                      & 0.314 & 0.488 & 0.650 \\
Non-agricultural countries                  & -0.416               & -0.206               & 0.006                &                      & 0.278 & 0.478 & 0.681 \\
Pre-1980 sample                             & -0.312               & 0.011                & 0.339                &                      & 0.593 & 0.710 & 0.829 \\
Post-1980 sample                            & -0.426               & -0.234               & -0.043               &                      & 0.604 & 0.723 & 0.844 \\
Inc. precipitations                         & -0.443               & -0.282               & -0.116               &                      & 0.393 & 0.520 & 0.655 \\
Inc. squared GDP                            & -0.482               & -0.290               & -0.098               &                      & 0.584 & 0.732 & 0.874 \\
Inc. squared temperatures                   & -0.501               & -0.310               & -0.115               &                      & 0.575 & 0.724 & 0.872 \\
Inc. squared volatilities                   & -0.957               & -0.693               & -0.444               &                      & 0.567 & 0.723 & 0.873 \\
Inc. squared volatilities, rich cts 	& -0.492               & -0.269               & -0.059               &                      & 0.224 & 0.393 & 0.561 \\
Region-Year FEs                             & -0.355               & -0.198               & -0.030               &                      & 0.670 & 0.794 & 0.917 \\
\midrule
CRU-TS dataset                            & -0.575               & -0.394                         & -0.210	& 			& 0.946 & 1.118 & 1.286                \\
\bottomrule
\end{tabular}
\caption{ \textsc{Short-term impact of an increase in temperature volatility.} \\ 
Contemporaneous responses of GDP levels and GDP volatilities to an exogenous 1$^\text{o}$C increase in the conditional volatility of annual temperatures ($h^T_{i,t}$). The responses are measured in the year when the shock takes place. The rows refer to alternative specifications of the panel VAR model: for each specification, the table reports the median responses along with the 16th and 84th percentiles of the posterior distribution. The sample includes 133 countries and it covers the 1961-2005 period, except for the CRU-TS dataset that runs to 2019.}              
\label{table_shortrun}            
\end{table}

\clearpage

\begin{table}[h!]
\begin{tabular}{lrrrrrrr}
\toprule
                                            & \multicolumn{3}{c}{GDP} 		& 						& \multicolumn{3}{c}{\ GDP volatility} 	\\
                                            & 16\%                 & 50\%                           & 84\%                 &   \ \ 	& 16\%                 & 50\%                               & 84\%                 \\
\midrule
Baseline                                    & -1.486               & -1.103               & -0.746               &                      & 0.419  & 0.616  & 0.822  \\
Poor countries                              & -1.618               & -1.144               & -0.643               &                      & 0.516  & 0.776  & 1.019  \\
Hot countries                               & -1.292               & -0.855               & -0.413               &                      & 0.442  & 0.669  & 0.896  \\
Rich countries                              & -0.874               & -0.483               & -0.075               &                      & -0.108 & 0.153  & 0.394  \\
Non-agricultural countries                  & -1.132               & -0.639               & -0.143               &                      & -0.195 & 0.091  & 0.383  \\
Pre-1980 sample                             & -0.952               & -0.519               & -0.086               &                      & 0.497  & 0.647  & 0.793  \\
Post-1980 sample                            & -1.031               & -0.600               & -0.179               &                      & 0.469  & 0.630  & 0.787  \\
Inc. precipitations                         & -0.237               & -0.105               & 0.028                &                      & 1.237  & 1.283  & 1.335  \\
Inc. squared GDP                            & -1.327               & -0.964               & -0.591               &                      & 0.402  & 0.600  & 0.795  \\
Inc. squared temperatures                   & -1.326               & -0.961               & -0.581               &                      & 0.384  & 0.590  & 0.781  \\
Inc. squared volatilities                   & -1.750               & -1.351               & -0.931               &                      & 0.338  & 0.547  & 0.744  \\
Inc. squared volatilities, rich cts 	& -0.984               & -0.504               & -0.048               &                      & -0.206 & 0.046  & 0.301  \\
Region-Year FEs                             & -0.627               & -0.323               & -0.015               &                      & 0.562  & 0.731  & 0.892  \\
\midrule
CRU-TS   dataset 			& -1.687 & -1.320 & -0.934               &                      & 0.945                & 1.157                & 1.358                \\
\bottomrule
\end{tabular}
\caption{ \textsc{Long-term impact of an increase in temperature volatility.} \\ 
Long-run response of GDP levels and GDP volatilities to an exogenous 1$^\text{o}$C increase in the conditional volatility of annual temperatures ($h^T_{i,t}$). The responses are measured 5 years after the materialization of the shock; the GDP response is calculated cumulating changes in growth rates over the 5-year horizon. The rows refer to alternative specification of the panel VAR model: for each specification, the table reports the median responses along with the 16th and 84th percentiles of the posterior distribution. The sample includes 133 countries and it covers the 1961-2005 period, except for the CRU-TS dataset that runs to 2019.}
\label{table_longrun}            
\end{table}

\clearpage \pagebreak


\clearpage
\pagebreak

\clearpage
\pagebreak

\clearpage
\pagebreak


\clearpage
\pagebreak

\pagebreak


\clearpage
\pagebreak


\clearpage
\pagebreak


\end{document}